# Early Accurate Results for Advanced Analytics on MapReduce

Nikolay Laptev
University of California,
Los Angeles
nlaptev@cs.ucla.edu

Kai Zeng
University of California,
Los Angeles
kzeng@cs.ucla.edu

Carlo Zaniolo
University of California,
Los Angeles
zaniolo@cs.ucla.edu

## ABSTRACT

Approximate results based on samples often provide the only way in which advanced analytical applications on very massive data sets can satisfy their time and resource constraints. Unfortunately, methods and tools for the computation of accurate early results are currently not supported in MapReduce-oriented systems although these are intended for 'big data'. Therefore, we proposed and implemented a non-parametric extension of Hadoop which allows the incremental computation of early results for arbitrary workflows, along with reliable on-line estimates of the degree of accuracy achieved so far in the computation. These estimates are based on a technique called bootstrapping that has been widely employed in statistics and can be applied to arbitrary functions and data distributions. In this paper, we describe our Early Accurate Result Library (EARL) for Hadoop that was designed to minimize the changes required to the MapReduce framework. Various tests of EARL of Hadoop are presented to characterize the frequent situations where EARL can provide major speed-ups over the current version of Hadoop.

## 1. INTRODUCTION

In today's fast-paced business environment, obtaining results quickly represents a key desideratum for 'Big Data Analytics' [16]. For most applications on large datasets performing careful sampling and computing early results from such samples provide a fast and effective way to obtain approximate results within the prescribed level of accuracy. Although the need for approximation techniques obviously grow with the size of the data sets, general methods and techniques for handling complex tasks are still lacking in both MapReduce systems and parallel databases although these claim 'big data' as their forte. Therefore in this paper, we focus on providing this much needed functionality. To achieve our goal, we explore and apply powerful methods and models developed in statistics to estimate results and the accuracy obtained from sampled data [27, 11]. We propose a method and a system that optimize the work-flow computation on massive data-sets to achieve the desired accuracy while minimizing the time and the resources required. Our approach is effective for analytical applications of arbitrary complexity and is supported by an Early Accurate Result Library ($EARL$) that we developed for Hadoop, which will be released for experimentation and non-commercial usage [1]. The early approximation techniques presented here are also important for fault-tolerance, when some nodes fail and error estimation is required to determine if node recovery is necessary.

The importance of EARL follows from the fact that real-life applications often have to deal with a tremendous amount of data. Performing analytics and delivering exact query results on such large volumes of stored data can be a lengthy process, which can be entirely unsatisfactory to a user. In general, overloaded systems and high delays are incompatible with a good user experience; moreover approximate answers that are accurate enough and generated quickly are often of much greater value to users than tardy exact results. The first line of research work pursuing similar objectives is that of Hellerstein et al. [15], where early results for simple aggregates are returned. In EARL however, we seek an approach that is applicable to complex analytics and dovetails with a MapReduce framework.

When computing some analytical function in EARL, a uniform sample, $s$, of stored data is taken, and the resulting error is estimated using the sample. If the error is too high, then another iteration is invoked where the sample size is expanded and the error is recomputed. Using sampling allows for a reduced computation and/or I/O costs for functions such as the *mean* and the *median*. This process is repeated until the computed error is below the user-defined threshold. The error for arbitrary analytical functions can be estimated via the bootstrapping technique described in [27]. This technique relies on resampling methods, where a number of samples are drawn from $s$. The function of interest is then computed on each sample producing a distribution used for estimating various accuracy measures of interest. Sampling in the bootstrapping technique is done *with replacement*, and therefore an element in the resample may appear more than once. There exist other resampling techniques, such as the jackknife [11], which perform resampling without replacement. The difference between the various resampling methods is in (1) the number of resamples required to obtain a reliable error estimate and in (2) the sampling method used. In this paper we use the bootstrapping technique because of its generality and accuracy [28].

While incorporating other resampling methods provides an exciting research direction for future work, it is beyond the limited scope of this paper.

Hadoop is a natural candidate for implementing EARL. In fact, while our early result approximation approach is general, it benefits from the fundamental Hadoop infrastructure. Hadoop employs a data *re-balancer* which distributes HDFS [2] data uniformly across the DataNodes in the cluster. Furthermore, in a MapReduce framework there are a set of (key, value) pairs which map to a particular reducer. This set of pairs can be distributed uniformly using random hashing and by choosing a subset of the keys at random, a uniform sample can be generated quickly. These two features make Hadoop a desirable foundation for *EARL*, while Hadoop's popularity maximizes the potential for practical applications of this new technology.

Thus, as an underlying query processing engine we chose Hadoop [2]. Hadoop, and more generally the MapReduce framework, was originally designed as a batch-oriented system, however it is often used in an interactive setting where a user waits for her task to complete before proceeding with the next step in the data-analysis workflow. With the introduction of high-level languages such as Pig [21], Sawzall [23] and Hive [30], this trend had accelerated. Due to its batch oriented computation mode, traditional Hadoop provides a poor support for interactive analysis. To overcome this limitation, Hadoop Online Prototype (HOP) [7] introduces a pipelined Hadoop variation in which a user is able to refine results interactively. In HOP however, the user is left with the responsibility of devising and implementing the accuracy estimation and improvement protocols. Furthermore in HOP, there is no feedback mechanism from the reducer back to the mapper, which is needed to effectively control the dynamically expanding sample.

Because *EARL* can deliver approximate results, it is also able to provide fault-tolerance in situations where there are node failures. Fault-tolerance is addressed in Hadoop via data-replication and task-restarts upon node failures, however with *EARL* it is possible to provide a result and an approximation guarantee despite node failures without task restarts.

Our approach, therefore, addresses the most pressing problem with Hadoop and with MapReduce framework in general: a high latency when processing large data-sets. Moreover, the problem of reserving too many resources to ensure fault-tolerance can also be mitigated by our approach and is discussed in Section 3.4.

## 1.1 Contributions and Organization

The paper makes the following three contributions:

1. A general early-approximation method is introduced to compute accurate approximate results with reliable error-bound estimation for arbitrary functions. The method can be used for processing large data-sets on many systems including Hadoop, Teradata, and others. An Early Accurate Result Library (EARL) was implemented for Hadoop and used for the experimental validation of the method.

2. An improved resampling technique was introduced for error estimation; the new technique uses delta maintenance to achieve much better performance.

3. A new sampling strategy is introduced that assures a more efficient drawing of random samples from a distributed file system.

**Organization**. In section 2 we describe the architecture of our library as it is implemented on Hadoop. Section 3 describes the statistical techniques used for early result approximation. Section 4 presents the resampling optimizations. In Sections 6 and 7 we empirically validate our findings and discuss related works that inspired some of our ideas. Finally, Section 8 draws conclusions about our current and future work.

## 2. ARCHITECTURE

This section describes the overall EARL architecture and gives a background on the underlying system. For a list of all symbols used refer to Table 1. EARL consists of (1) a sampling stage, (2) a user's task, and (3) an accuracy estimation stage which are presented in Figure 1. The sampling stage draws a uniform sample $s$ of size $n$ from the original data set $S$ of size $N$ where $n << N$. In Section 3.3 we discuss how this sampling is implemented using tuple-based and key-based sampling for MapReduce. After the initial sample $s$ is drawn from the original data-set, $B$ samples (i.e. resamples) with replacement are taken from $s$. These resamples are used in the work phase (user's task) to generate $B$ results, which are then used to derive a result distribution [9] in the accuracy estimation phase. The sample result distribution is used for estimating the accuracy. If the accuracy obtained is unsatisfactory, the above process is repeated by drawing another sample $\Delta s$ which is aggregated with the previous sample $s$ to make a larger sample $s'$ for higher accuracy. The final result is returned when a desired accuracy is reached.

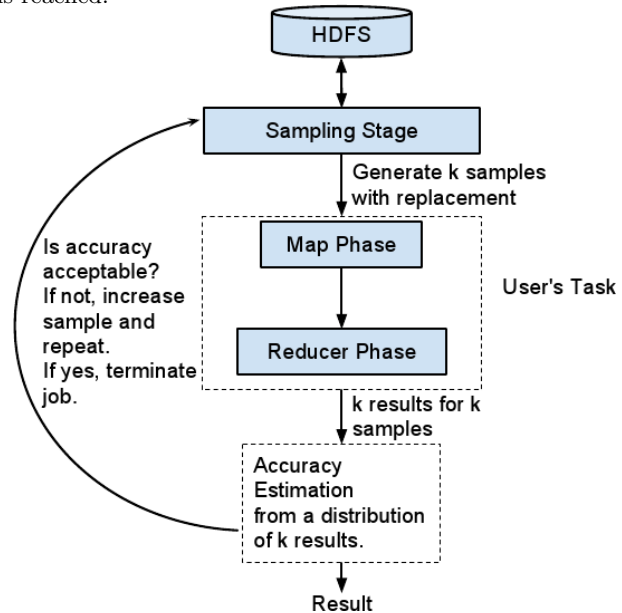

**Figure 1: A simplified EARL architecture**

## 2.1 Extending MapReduce

The MapReduce (MR) model is becoming increasingly popular for tasks involving large data processing. The programming model adopted by MapReduce was originally inspired by functional programming. In the MR model two

1029

| Symbol | Description |
|---|---|
| $B$ | Number of bootstraps |
| $b$ | A particular bootstrap sample |
| $n$ | Sample size |
| $s$ | Array containing the sample |
| $p$ | Percentage of the data contained in a sample |
| $N$ | Total data size |
| $S$ | Original data-set |
| $F_i$ | File split $i$ |
| $c_v$ | Coefficient of variation |
| $f$ | Statistic of interest |
| $\sigma$ | User desired error bound |
| $\tau$ | Error accuracy |
| $X_i$ | A particular data-item $i$ |
| $D$ | Total amount of data processed |

Table 1: Symbols used

main *stages*, map and reduce, are defined with the following signatures:

$$map : (k_1, v_1) \to (k_2, list(v_2))$$

$$reduce : (k_2, list(v_2)) \to (k_3, v_3)$$

The map function is applied on every tuple $(k_1, v_1)$ and produces a list of intermediate $(k_2, v_2)$ pairs. The reduce function is applied to all intermediate tuples with the same key producing $(k_3, v_3)$ as output. The MapReduce model makes it simple to parallelize *EARL*'s approximation technique introduced in Section 3.1.

Hadoop, an open source implementation of the MapReduce framework, leverages Hadoop Distributed File System (HDFS) for distributed storage. HDFS stores file system metadata and application data separately. HDFS stores metadata on a dedicated node, termed the *NameNode* (other systems, such as the Google File System (GFS) [13] do likewise). The application data is stored on servers termed *DataNodes*. All communication between the servers is done via TCP protocols. The file block-partitioning, the replication, and the logical data-splitting provided by HDFS simplify *EARL*'s sampling technique, as discussed in Section 3.3.

For implementing the underlying execution engine, we evaluated 3 alternatives, (1) Hadoop, (2) HaLoop [4] and (3) Hadoop online [7]. Although HaLoop would allow us to easily expand the sample size on each iteration, it would be slow for non-iterative MR jobs due to the extra overhead introduced by HaLoop. With Hadoop online, we would get the benefit of pipelining, however further modifications would be needed to allow the mapper to adjust the current sample size. Since both Hadoop Online and HaLoop do not exactly fit our requirements, we therefore decided to make a relatively simple change to Hadoop that would allow dynamic input size expansion required by out approach. Thus *EARL* adds a simple extension to Hadoop to support dynamic input and efficient resampling. An interesting future direction is to combine *EARL*'s extensions with those of HaLoop and HOP to make a comprehensive data-mining platform for analyzing massive data-sets. In summary, with the goals of seeking *EARL* fast and requiring the least amount of changes to the core Hadoop implementation we decided to use the default version of Hadoop instead of using Hadoop extensions such as Hadoop online or HaLoop .

To achieve dynamic input expansion we modify Hadoop in three ways (1) to allow the reducers to process input before mappers finish (2) to keep mappers active until explicitly terminated and (3) to provide a communication layer between the mappers and reducers for checking the termination condition. While the first goal is similar to that of pipelining implemented in Hadoop Online Prototype (HOP) [7], *EARL* is different from HOP in that in *EARL* the mapper is actively, rather than a passively, transfers the input to the reducer. In other words, the mapper actively monitors the sample error and actively expands the current sample size. The second goal is to minimize the overall execution time, thus instead of restarting a mapper every time sample size expands, we reuse an already active mapper. Finally, each mapper monitors the current approximation error and is terminated when the required accuracy is reached.

We also modify the reduce phase in Hadoop to support efficient incremental computation of the user's job. We extend the MapReduce framework with a finer-grained reduce function, to implement incremental processing via four methods: (i) *initialize()*, (ii) *update()*, (iii) *finalize()* and (iv) *correct()*. The *initialize*() function reduces a set of data values into a *state*, i.e. $<k, v_1>, <k, v_2>, ..., <k, v_k> \to <k, state>$. A state is a representation of a user's function $f$ after processing $s$ on $f$. Each resample will produce a state. Saving states instead of the original data requires much less memory as needed for fast in-memory processing. The *update*() function updates the state with a new input which can be another state or a <key, value> pair. The *finalize*() function computes the current error and outputs the final result. The *correct*() function takes the output of the *finalize*() function, and corrects the final result. When computed from a subset of the original data, some user's tasks need to be adjusted in order to get the right answer. For example, consider a SUM query which sums all the input values. If we only use $p$ of the input data, we need to scale the result by $1/p$. As the system is unaware of the internal semantics of user's MR task, we allow our users to specify their own correction logic in *correct*() with a system provided parameter $p$ which is the percentage of the data used in computation.

Hadoop's limited two stage model makes it difficult to design advanced data-mining applications for which reason high level languages such as PIG [21] were introduced. *EARL* does not change the logic of the user's MapReduce programs and achieves the early result approximation functionality with minimal modifications to the user's MR job (see Figure 4). Next the accuracy estimation stage is described.

## 3. ESTIMATING ACCURACY

In EARL, error estimation of an arbitrary function can be done via resampling. By re-computing a function of interest many times, a result distribution is derived from which both the approximate answer and the corresponding error are retrieved. EARL uses a clever delta maintenance strategy that dramatically decreases the overhead of computation. As a measurement of error, in our experiments, we use a coefficient of variation ($c_v$) which is a ratio between the standard deviation and the mean. Our approach is independent of the error measure and is applicable to other errors (e.g., bias, variance). Next a traditional approach to error estimation is presented, after which our technique is discussed.

A crucial step in statistical analysis is to use the given data to estimate the accuracy measure, such as the bias, of a given statistic. In a traditional approach, the accuracy measure is computed via an empirical analog of the explicit theoretical

1030

formula derived from a postulated model [27]. Using variance as an illustration let $X_1, ..., X_n$ denote the data set of $n$ independent and identically distributed (i.i.d.) data-items from an unknown distribution $F$ and let $f_n(X_1, ..., X_n)$ be the function of interest we want to compute. The variance of $f_n$ is then:

$$var(f_n) = \int [f_n(x) - \int f_n(y) d\prod_{i=1}^{n} F(y_i)]^2 d\prod_{i=1}^{n} F(x_i) \quad (1)$$

where $x = (x_1, ..., x_n)$ and $y = (y_1, ..., y_n)$. Given a simple $f_n$ we can obtain an equation of $var(f_n)$ as a function of some unknown quantities and then substitute the estimates of the unknown quantities to estimate the $var(f_n)$. In the case of the sample mean, where $f_n = \bar{X}_n = n^{-1} \sum_{i=1}^{n} X_i$, $var(\bar{X}_n) = n^{-1} var(X_1)$. We can therefore estimate $var(\bar{X}_n)$ by estimating $var(X_1)$ which is usually estimated by the sample variance $(n-1)^{-1} \sum_{i=1}^{n} (X_i - \bar{X}_n)^2$. The use of Equation 1 to estimate the variance is computationally feasible only for simple functions, such as the mean. Next we discuss a resampling method used to estimate the variance of arbitrary functions.

Resampling approaches, such as *bootstrap* and *jackknife* [29], provide an accuracy estimation for general functions. Both of these resampling methods do not require a theoretical formula to produce the error estimate of a function. In fact these techniques allow for estimation of the sampling distribution of almost any statistic using only very simple methods [12]. The estimate of the variance can then be determined from the sampling distribution. Both techniques however require repeated computation of the function of interest on different resamples. The estimate of the variance of the result, $\sigma$, produced by this repeated computation is $\sigma^2(F) = E_F(\hat{\theta} - E_F(\hat{\theta}))^2$, where $\theta$ is the parameter of interest. The jackknife has a fixed requirement for the number of resamples, $n$, that is often relatively low. The number of samples required by the *bootstrap* approach, however, is not fixed and can be much lower than that of the *jackknife*. Moreover, it is known that jackknife does not work for many functions such as the median [11]. Thus, in this first version of EARL we concentrate on bootstrapping and leave the study of other resampling methods for future work.

To compute an *exact* bootstrap variance estimate $\binom{2n-1}{n-1}$ resamples are required, which for $n = 15$ is already equal to $77 \times 10^6$, therefore an approximation is necessary to make the bootstrap technique feasible. The *Monte-Carlo* [27] is the standard approximation technique used for resampling methods including the bootstrap that requires less than $n$ resamples. It works by taking $B$ resamples resulting in variance estimate of $\hat{\sigma}_B^2 = \frac{1}{B} \sum_{n=1}^{B} (\hat{\theta}_n^* - \hat{\theta}^*)^2$ where $\hat{\theta}^*$ is the average of $\hat{\theta}_n^*$'s. The theory suggests that $B$ should be set to $\frac{1}{2} \epsilon_0^{-2}$ [12], where $\epsilon_0$ corresponds to the desired error of the Monte Carlo approximation with respect to the the original bootstrap estimator. Experiments, however, show that a much better value of $B$ can be used in practical applications, therefore in Section 3.2 we develop an algorithm to empirically determine a good value of $B$.

### 3.1 Accuracy Estimation Stage

The accuracy estimation stage ($AES$) uses the bootstrap resampling technique [9] outlined in the previous subsection to estimate the standard error $c_v$ of the statistic $f$ computed from sample $s$.

In many applications, the number of bootstrap samples required to estimate $c_v$ to within a desired accuracy $\tau$ can be substantial. $\tau$ is defined as $\tau = (c_{v_i} - c_{v_{i+1}})$ which measures the stability of the error. Before performing the approximation, we estimate the required $B$ and $n$ to compute $f$ with $c_v \leq \sigma$. If $B \times n \geq N$, then EARL informs the user that an early estimation with the specified accuracy is not faster than computing $f$ over $N$ and instead the computation over the entire data-set is performed. AES allows for error estimation of general MR-Jobs (mining algorithms, complex functions etc). Furthermore, EARL works on categorical, as well as on inter-dependent data as discussed in Appendix A.

For completeness, we will first discuss how $B$ and $n$ impact the error individually, and then in Section 3.2 we present an algorithm to pick $B$ and $n$ that empirically minimizes the product $B \times n$. Figure 2 (left) shows how $B$ affects $c_v$ experimentally. Normally roughly 30 bootstraps are required to provide a confident estimate of the error. The sample size, $n$, given a fixed $B$ has a similar effect on $c_v$ as shown in Figure 2 (right). A larger $n$ results in a lower error. Depending on the desired accuracy, $n$ can be chosen appropriately as described next.

### 3.2 Sample Size and Number of Bootstraps

To perform resampling efficiently (i.e. without processing more data than is required) we need to minimize the sample size ($n$) and the number of resamples performed ($B$). A straightforward sample size adjustment might work as follows: pick an initial sample size $s$ of size $n$ which theoretically achieves the desired error $\sigma$ and compute $f$ on $s$. If the resulting error $\hat{\sigma}$ is greater than $\sigma$ then the sample size is increased (e.g., doubled). A similar naïve strategy may be applicable when estimating the minimum $B$. This naïve solution however may result in an overestimate of the sample size and the number of resamples. Instead, following [5] we propose a two phase algorithm to estimate the final early approximate result satisfying the desired error bound while empirically minimizing $B \times n$. As shown later, our algorithm requires only a single iteration.

*Sample Size And Bootstrap Estimation* (SSABE) algorithm performs we propose, performs the following operations: (1) In the first phase, it estimates the minimum $B$ and $n$ and then (2) in the second phase, it evaluates the function of interest, $f$, $B$ times on $s$ of size $n$. To estimate the required $B$, the first phase an initially small $n$, a fraction $p$ of $N$, is picked. In practice we found that $p = 0.01$ gives robust results. Given a fixed $n$, a sample $s$ is picked. The function $f$ is then computed for different candidate values of $B$ ($\{2, ..., \frac{1}{\tau}\}$). The execution terminates when the difference $|c_{v_i} - c_{v_{i-1}}| < \tau$. The $B$ value so determined is used as the estimated number of bootstraps. In practice the value of $B$ so calculated is much smaller than the theoretically predicted $\frac{1}{2} \epsilon_0^{-2}$.

To estimate the required sample size $n$, first the initial sample size $\frac{1}{\tau}$ is picked. The initial sample is split into $l$ smaller subsamples $s_i$ each of size $n_i$ where $n_i = \frac{n}{2^{l-i}}$ and $1 \leq i \leq l$. In our experiments we found it to be sufficient to set $l = 5$. For each $s_i$ we compute the $c_v$ using $B$ resamples. When computing $f$ on $s_i$ we perform delta maintenance discussed in Section 4. The result will be a set of points $A[s_i] = c_v$. For these set of points, the best fitting curve is constructed. The curve fitting is done using the standard method of least squares. The best fitted curve yields an $s_i$ that satisfies the given $\sigma$. Finally, once the estimate for $B$ and $n$ is complete, the second phase is invoked where the actual user job is executed using $s$ of size $n$ and $B$.

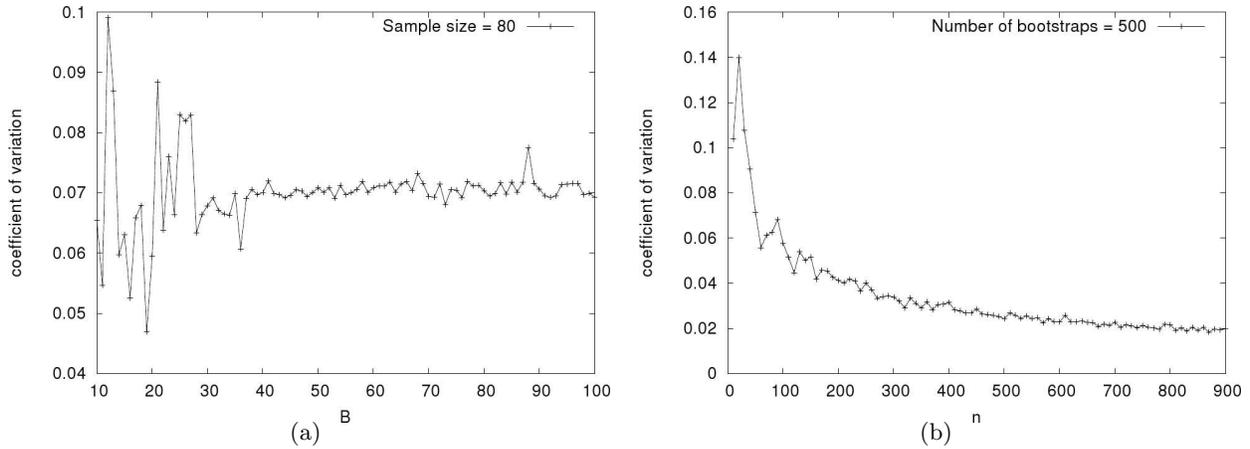

Figure 2: (a) Effect of $B$ on $c_v$, (b) Effect of $n$ on $c_v$

The initial $n$ is picked to be small, therefore the sample size and the number of bootstraps estimation can be performed on a single machine prior to MR job start-up. Thus, when performing the estimation for $n$ and $B$ we run the user's MR job in a local mode without launching a separate JVM. Using the local-mode we avoid running the mapper and the reducer as separate JVM tasks and instead a single JVM is used which allows for a fast estimation of the required parameters needed to start the job.

### 3.3 Sampling

In order to provide a uniformly random subset of the original data-set, *EARL* performs sampling. While sampling over memory-resident, and even disk resident, data had been studied extensively, sampling over a distributed file system, such as HDFS, has not been fully addressed [20]. Therefore, we provide two sampling techniques: (1) pre-map sampling and (2) post-map sampling. Each of the techniques has its own strengths and weaknesses as discussed next.

In HDFS, a file is divided into a set of blocks, each block is typically 64MB. When running an MR job, these blocks can be further subdivided into "Input Splits" which are used as input to the mappers. Given such an architecture, a naïve sampling solution is to pick a set of blocks $B_i$ at random, possibly splitting $B_i$ into smaller splits, to satisfy the required sample size. This strategy however will not produce a uniformly random sample because each of the $B_i$ and each of the splits can contain dependencies (e.g., consider the case where data is clustered on a particular attribute resulting in clustered items to be placed next to each other on disk due to spatial locality). Another naïve solution is to use a reservoir smapling algorithm to select $n$ random items from the original data-set. This approach produces a uniformly random sample, but it suffers from slow loading times because the entire dataset needs to be read, and possibly re-read when further samples are required. We thus seek a sampling algorithm that avoids such problems.

In a MapReduce environment, sampling can be done before or while sending the input to the Mapper (*pre-map* and *post-map* sampling respectively). Pre-map sampling significantly reduces the load times, however the sample produced may be an inaccurate representation of the total $<k,v>$ pairs present in the input. Post-map sampling first reads the data and then outputs a uniformly random sample of desired size. Post-map sampling also avoids the problem of inaccurate $<k,v>$ counts.

*Post-map* sampling works by reading and parsing the data before sending the selected $<k,v>$ pairs to the reducer. Each $<k,v>$ pair is stored by using random hashing that generates a pre-determined set of keys, of size proportional to the required sample size. We store all *key, value* pairs on the mapper locally, and when all data had been received, we randomly pick $p$ *key, value* pairs that satisfies the sample size and send it to the reducer. Because sampling is done without replacement, the key, value pairs already sent are removed from the hashmap. Post-map sampling is shown in Algorithm 1.

---

**Algorithm 1** Post-map sampling

$hash \leftarrow$ initialize the hash
2: $timestamp \leftarrow$ initialize the timestamp
**while** input $!=$ null **do**
4:   $key \leftarrow$ get random key for input
  $value \leftarrow$ get value for input
6:   $hash[key] \leftarrow value$
**end while**
8: sendSample(hash(rand()%hash_size)
**while** true **do**
10:   **if** get_new_error_average (timestamp) $>$ required **then**
    sendSample(hash(rand()%hash_size))
12:   **else**
    return
14:   **end if**
**end while**

---

Unlike *post-map* sampling, which first reads the entire dataset and then randomly chooses the required subset to process, *pre-map* sampling works by sampling a portion $p$ of the initial dataset before it gets passed into the mapper. Therefore, because sampling is done prior to data loading stage, the response time is greatly improved, with a potential downside of a slightly less accurate result. The reason for this is because when sampling from HDFS directly, we can efficiently only do so by sampling lines[1]. Each line however may contain a variable number of $<k, v>$ pairs so that when producing a 1% sample of the *key,value* pairs, we may produce a larger or a lesser sample. Therefore, for $f$ which needs correction, we would be unable to do so accurately without additional information from the user. For majority of the cases however correcting the final result is not necessary, and even for cases when correction is required, the estimate of the number of the *key, value* pairs produced by the *pre-map* sampling approach is good enough in practice. Nevertheless the user has the flexibility to use *post-map* sampling if an accurate correction to the final result is desired.

We assume, w.l.o.g., that the input is delimited by *newlines*, as opposed to commas or other delimiters. A set of logical splits is first generated from the original file which will be used for sampling. For each split $F_i$, we maintain a bit-vector representing the *start* byte locations of the lines we had already included in our sample. Therefore until the required sample size is met, we continue picking a random $F_i$ and a random *start* location which will be used to include a line from a file. To avoid the problem of picking a file *start* location which is not a beginning of a line, we use Hadoop's *LineRecordReader* to backtrack to the beginning of a line. Using *pre-map* sampling we avoid sending an overly large amount of data to the mapper which improves response time as seen in experiment in Section 6.1. In rare cases where a larger sample size is required for an in-progress task, a new split is generated and the corresponding map task is restarted in the *TaskInProgress* Hadoop class. Algorithm 2 presents the HDFS sampling algorithm used in pre-map sampling.

---

**Algorithm 2** HDFS sampling algorithm used in pre-map sampling

$start \leftarrow split.getStart()$
$end \leftarrow start + split.getLength()$
$sample \leftarrow \emptyset$
**while** $|sample| < n$ **do**
  $start \leftarrow$ pick a random start position
  **if** start $! =$ beginning of a line **then**
    skipFirstLine $\leftarrow$ true
    fileIn.seek(start)
  **end if**
  in $=$ new LineReader(fileIn, job)
  **if** skipFirstLine **then**
    start $+=$ in.readLine(new Text(),
    0, (int)Math.min((long)Integer.MAX_VALUE,
    end - start))
  **end if**
  sample $\leftarrow$ includeLineInSample()
  skipFirstLine $\leftarrow$ false
**end while**

---

Therefore, while pre-map is fast and works well for most cases, post-map is still very useful for applications where a correction function relies on an accurate estimate of the total *key, value* pairs. Experiments highlighting the difference between the two sampling methods are presented in experiment of Section 6.5.

---

[1] A default file format in Hadoop is a line delimitted by a new-line character. Another format can be specified via the *RecordReader* class in Hadoop.

In both the post-map and the pre-map sampling, every reducer writes its computed error together with a time-stamp onto HDFS. These files are then read by the mappers to compute the overall average error. Because both the mappers and the reducers share the same JobID, it is straight forward to list all files generated by the reducers within the current job. The mapper stores a time-stamp that corresponds to the last successful read attempt of the reducer output. The mapper collects all errors, and computes the average error. The average error, incurred by all the reducers, is used to decide if sample size expansion is required. Lines 9-15 in Algorithm 1 demonstrate this for pos-map sampling.

Note that in a MapReduce framework independence is assumed between *key, value* pairs. In addition to being natural in a MapReduce environment, the independence assumption also makes sampling applicable to algorithms relying on capturing data-structure such as correlation analysis.

### 3.4 Fault Tolerance

Most clusters that use Hadoop and the MapReduce frameworks utilize commodity hardware and therefore node failure are a part of every-day cluster maintenance. Node failure is handled in the Hadoop framework with the help of data-replication and task-restarts upon failures. Such practices however can be avoided if the user is only interested in an approximate result. Authors in [26] show that in the real world, over 3% of hard-disks fail per year, which means that in a server farm with 1,000,000 storage devices, over 83 will fail every day. Currently, the failed nodes have to be manually replaced, and the failed tasks have to be restarted. Given a user specified approximation bound however, even when most of the nodes have been lost, a reasonable result can still be provided. Using the ideas from *AES* stage the error bound of the result can still be computed with a reasonable confidence. Using our simple framework, a system can therefore be made more robust against node failures by delivering results with an estimated accuracy despite node failures.

## 4. OPTIMIZATIONS

The most computationally intensive part of *EARL*, aside from the user's job $j$, is the re-execution of $j$ on an increasingly larger sample sizes, during both the main job execution and during initial sample size estimation. One important observation is that this intensive computation can reuse its results from the previous iterations. By utilizing this incremental processing, performing large-scale computations can be dramatically improved. We first take a more detailed look at the processing of two consecutive bootstrap iterations and then we discuss the optimization of the bootstrapping (resampling) procedure so that when recomputing $f$ on a new resample $s'$ we can perform delta maintenance using a previous resample $s$.

### 4.1 Inter-Iteration Optimization

Let $s$ denote the sample of size $n$ used in the $i$-th iteration, and $\{b_i, 1 \leq i \leq B\}$ denote the $B$ bootstrap resamples drawn from $s$. The user's job $j$ is repeated on all $b_i$'s. In the $(i+1)$-th iteration, we enlarge sample $s$ with another sample $\Delta s$. $s$ and $\Delta s$ are combined to get a new sample $s'$ of size $n'$. $B$ bootstrapping resamples $\{b'_i, 1 \leq i \leq B\}$ are drawn from $s'$, and the user's job $j$ is repeated on all $b'_i$'s. Each resample $b'_i$ can be decomposed into two parts: (1) the set of data-items randomly sampled from $s$, denoted by $b'_{i,s}$, and (2) the set of data-items randomly sampled from $\Delta s$, denoted by $b'_{i,\Delta s}$.

Therefore, in the $(i+1)$-th iteration, instead of drawing a completely new $\{b'_i\}$ from $s'$, we can reuse the resamples $\{b_i\}$ generated in the $i$-th iteration. The idea is to generate $b'_{i,s}$ by updating $b_i$, and to generate $b'_{i,\Delta s}$ by randomly sampling from $\Delta s$. This incremental technique has two benefits, in that we can save a part of: (1) the cost of bootstrapping resampling $\{b'_i\}$, and (2) the computation cost of repeating the user's job $j$ on $\{b'_i\}$.

The process of generating $b'_{i,s}$ from $b_i$ is not trivial, due to the following observation. Each data item in $b'_i$ is drawn from $b_i$ with probability $\frac{n}{n'}$, and from $\Delta s$ with probability $1 - \frac{n}{n'}$. We have the following equation modeling the size of $b'_{i,s}$ by a binomial distribution.

$$P(|b'_{i,s}| = k) = \binom{n'}{k}\left(\frac{n}{n'}\right)^k \left(1 - \frac{n}{n'}\right)^{n'-k} \quad (2)$$

This means that we may need to randomly delete data-items from $b_i$, or add data-items randomly drawn from $s$ to $b_i$. We first present a naive algorithm which maintains a resample $b'_i$ from $s'$ by updating the resample $b_i$ form $s$ in three steps: (1) randomly generate $|b'_{i,s}|$ according to Equation 2. (2) if $|b'_{i,s}| < n$, then randomly delete $(n - |b'_{i,s}|)$ data-items from $b_i$; if $|b'_{i,s}| > n$, then randomly sample $(|b'_{i,s}| - n)$ data-items from $s$ and combine them with $b_i$. (3) generate $(n' - |b'_{i,s}|)$ random sample from $\Delta s$ and combine them with $b_i$.

The above process requires us to record all the data-items of $s$ and $b_i$, which is a huge amount of data that cannot reside in memory. Therefore, $s$ and $b_i$ must be stored on the HDFS file system. Because this data will be accessed frequently, the disk I/O cost can be a major performance bottleneck.

Next, we present our optimization algorithm with a cache mechanism that supports fast incremental maintenance. Our approach is based on an interesting observation from Equation 2. With $n'$ very large and $n/n'$ fixed, which is usually the case in massive MapReduce tasks, Equation 2 can be approximated by the Gaussian distribution

$$N\left(n, n\left(1 - \frac{n}{n'}\right)\right) \quad (3)$$

For a Gaussian distribution, by the famous *3-sigma rule*, most data concentrate around the mean value, to be specific, within 3 standard deviations of the mean. As an example, for the distribution 3 with its standard deviation denoted by $\sigma_0 = \sqrt{n\left(1 - \frac{n}{n'}\right)}$, over 99.7% data lie within the range $(n - 3\sigma_0, n + 3\sigma_0)$; over 99.9999% data lie within the range $(n - 5\sigma_0, n + 5\sigma_0)$. Note that $\sigma_0 < \sqrt{n}$.

Next we explain our optimized algorithm in more detail. For the $i$-th iteration, we define the delta sample added to the previous sample as $\Delta s_i$. For the first iteration, we can treat the initial sample as a delta sample added to an empty set. Therefore we can denote it by $\Delta s_1$. The size of $\Delta s_i$ is $n_i$. After the $i$-th iteration, a bootstrapping resample $b$ can be partitioned into $\{b_{\Delta s_k}, k < i\}$, where $b_{\Delta s_k}$ represents the data-items in $b$ drawn from $\Delta s_k$. We build a two-layer memory-disk structure of $b$. Instead of simply storing $b$ on a hard-disk, we build two pieces of information of it: (i) memory-layer information (a sketch structure) and (ii) disk-layer information (the whole data set). A *sketch* of data set of size $n$ is $c\sqrt{n}$ data items randomly drawn without replacement from it where $c$ is a chosen constant.

Determining an appropriate $c$ is a trade-off between memory space and the computation time. A larger $c$ will cost more memory space but will introduce less randomized update latency. The sketch structure contains $\{sketch(b_{\Delta s_k})\}$ and $\{sketch(\Delta s_k)\}$.

During updating, instead of accessing $s$ and $b$ directly, we always access the sketches first. Specifically, for step 2 in our algorithm, if we need to randomly delete data-items from $b_{\Delta s_k}$, we sequentially pick the data-items from $sketch(b_{\Delta s_k})$ for deletion; if we need to add data-items randomly drawn from $\Delta s_k$, we sequentially pick the data-items from $sketch(\Delta s_k)$ for addition. For already picked data-items, we mark them as *used*. At the end of each iteration, we will randomly substitute some of the unused data items in $sketch(b_{\Delta s_k})$ with the used data items in $sketch(\Delta s_k)$ by following a reservoir sampling approach, in order to maintain $sketch(b_{\Delta s_k})$ as a random sketch of $b_{\Delta s_k}$. If we use up all the data-items in a sketch, we access the copy stored in HDFS, applying two operations: (1) committing the changes on the sketch, and (2) resampling a new sketch from the data.

### 4.2 Intra-Iteration Optimization

When performing a resample, at times a large portion of the new sample is identical to the previous sample in which case effective delta maintenance can be performed. Our main observation is shown in Equation 4. The equation represents the probability that a fraction $y$ of a resample is identical to that of another resample. Therefore, for example if $n = 29$ and $y = 0.3$, that means that 35% of the time, resamples will contain 30% of identical data. In other words, for roughly 1 in 3 resamples, 30% of each resample will be identical to one-another. Because of the relatively high level of similarity among samples, an intra iteration delta maintenance can be performed where the part that is shared between the resamples is reused.

$$P(X = y) = \frac{n!}{(n - y*n)! \times n^{y*n}} \quad (4)$$

Using Equation 4 we can find the optimal $y$, given $n$, that minimizes the overall work performed by the bootstrapping stage. The overall work saved can be stated as $P(X = y)*y$. Figure 3 shows how the overall work saved varies with $n$ for different $y$. The optimal $y$ for given $n$ can be found using a simple binary search. Overall we found that on average we save over 20% of work using our Intra Iteration Optimization procedure when compared to the standard bootstrapping method.

While the optimization techniques presented in this section greatly increase the performance of the standard bootstrap procedure there is still more research to be done with regards to delta maintenance and sampling techniques. Our optimization techniques are best suited for small sample sizes, which is reasonable for a distributed system where both response time and data-movement must be minimized. Next, several implementation challenges that we faced are outlined including the overhead of restarting a MapReduce job.

### 5. CURRENT IMPLEMENTATION

We have used Hadoop version 0.20.2-dev, to implement our extension and run the experiments on a small cluster of machines of size 5 containing Intel Core duo (CPU E8400 @ 3.00GHz), 320MB of RAM and Ubuntu 11.0 32bit. Each of the parts shown in Figure 1 are implemented as separate modules which can seamlessly integrate with user's Map-Reduce jobs. The sampler is implemented by modifying the *RecordReader* class to implement the *Pre-Map sampling* and extending the *map* class to implement the *Post-Map sampling*. The resampling and update strategies are implemented by extending the *Reduce* class. The results generated from resamples are used for result and accuracy estimation in the *AES* phase. The *AES* phase computes the coefficient of variation ($c_v$) and outputs the result to HDFS which is read by the main Map-Reduce job where the termination condition is checked. Because the number of required resamples and the required sample size are estimated via regression, a single iteration is usually required. Figure 4 shows an example of an MR program written using EARL's API. As can be observed from the figure, the implementation allows for a generic *user_job* to take advantage of our early approximation library.

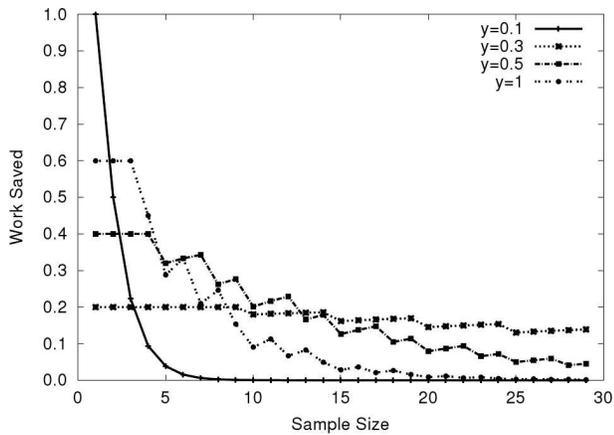

**Figure 3: Work saved using our intra iteration optimization vs. sample size**

The biggest implementation challenge with *EARL* was reducing the overhead of the *AES* phase and of the sample generation phase. If implemented naively, (i.e. making both the sampler and the *AES* phase its separate job) then the execution time would be inferior to that of the standard Hadoop especially for small data-sets and light aggregates where *EARL*'s early approximation framework has little impact to begin with. We wanted to make *EARL* light-weight so that even for light tasks, *EARL* would not add additional overhead and the execution time in the worst case would be comparable to that of the standard Hadoop.

The potential overhead of our system is due to three factors: (1) creating a new MR job for each iteration used for sample size expansion (2) generating a sample of the original dataset and (3) creating numerous resamples to compute a result distribution that will be used for error estimation. The first overhead factor is addressed with the help of pipelining, similar to that of Hadoop Online, however in our case the mappers also communicate with reducers to receive events that signal sample size expansion or termination. With the help of pipelining and efficient inter task communication, we are able to reuse Hadoop tasks while refining the sample size. The second challenge is addressed via the added feature of the mappers to directly ask for more splits to be assigned, in the case of pre-map sampling, when a larger sample size is required. Alternatively a sample can be generated using post-map sampling as discussed

```
public static void main(String[] args)
  throws Exception {
// Initialization of local variables
// ...
// ...
Sampler s = new Sampler();

while (error > sigma) {
 // path_string is the initial DataSet
 s.Init(path_string);
 // num_resamples of resamples of size
 // sample_size is generated. Both of
 // these variables are determined
 // empirically.
 s.GenerateSamples(sample_size,
                   num_resamples);
 JobConf aes_job =
   new JobConf(AES.class);

 JobConf user_job;

 // For each sample we execute user_job
 for (int i = 0; i < num_resamples; i++)
 {
  user_job = new JobConf(
    MeanBootstrap.class);
  // Init of the user_job
  // ...
  // ...
  JobClient.runJob(user_job);
 }

 // AES uses the input from user_job to
 // compute the
 // approximation error.
 // Init of the aes_job
 // ...
 // ...
 // The aes job also updates the err.
 JobClient.runJob(aes_job);
 // In cases where early approximation
 // is not possible, sample_size and
 // num_resamples will be set to N and 1
 // respectively.
 UpdateSampleSizeAndNumResamples();
 }
}
```

**Figure 4: An example of how a user_job would work with the EARL framework**

in Section 3.3. Post and pre-map sampling work flawlessly with the Hadoop infrastructure to deliver high quality random sample of the input data. Finally the last challenge is addressed via a resampling algorithm and its optimizations presented in Section 3.3. Re-sampling is actually implemented within a reduce phase, to minimize any overhead due to job restarts. Due to delta maintenance, introduced in Section 4, resampling becomes efficient and its overhead is tremendously decreased making our approach not only feasible but to deliver an impressive speed-up over standard Hadoop. Next key experiments are presented which showcase the performance of *EARL*.

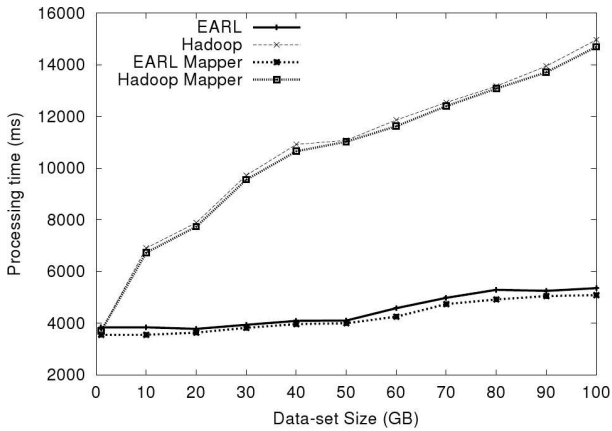

Figure 5: Computation of average using EARL and stock Hadoop

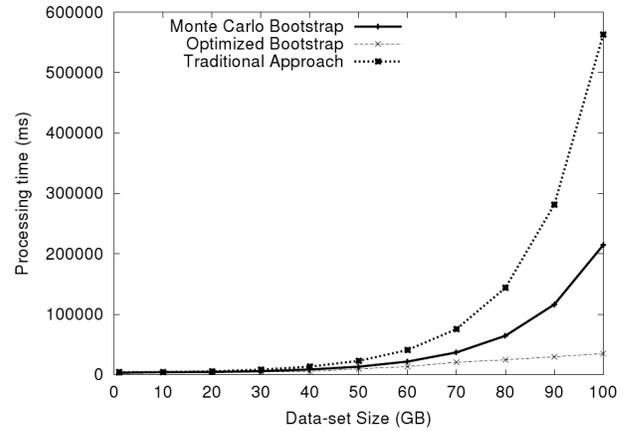

Figure 6: Computation of median using EARL and stock Hadoop

## 6. EXPERIMENTS

This section is aimed at testing the efficiency of our approach. A set of experiments measuring the efficiency of the accuracy estimation and sampling stages are presented in the following sections. To measure the asymptotic behavior of our approach a synthetically generated data-set is used. The synthetic dataset allows us to easily validate the accuracy measure produced by EARL. More experiments with advanced data-minng algorithms in real-world workflows is currently under way. The normalized error used for the approximate early answer delivery is 0.05 (i.e. our results are accurate to within 5% of the true answer). The experiments were executed on simple, single phase MR tasks to give concrete evidence of applicability of EARL, and more elaborate experiments on a wider range of mining algorithms is part of our future work.

### 6.1 A Strong Case for EARL

In this experiment, we implemented a simple MR task computing the mean, and tested it on both standard Hadoop and EARL. Figure 5 shows the performance comparison between these two. It shows that when the data-set size is relatively large ($> 100GB$), our solution provides an impressive performance gain (4x speed-up) over standard Hadoop even for a simple function such as the mean. In the worst case scenario, where our framework cannot provide early approximate results ($< 1GB$), our platform intelligently switch back to the original work flow which runs on the entire data-set without incurring a big overhead. It demonstrates clearly that EARL greatly outperforms the standard Hadoop implementation even for light-weight functions. Figure 5 also shows that a standard Hadoop data loading approach is much less efficient than the proposed pre-map sampling technique.

### 6.2 Approximate Median Computation

In this experiment, we did a break-down study, to measure how much a user defined MR task can benefit from resampling techniques and our optimization techniques. We used the computation of a median as an example, and tested it using three different implementations: (1) standard Hadoop, (2) original resampling algorithm, and (3) optimized resampling algorithm. Figure 6 shows that: (1) With a naïve Monte Carlo bootstrap, we can provide a reliable estimate for median with a 3-fold speed-up, compared to the standard Hadoop, due to a much smaller sample size requirement. (2) Our optimized algorithm provides another 4x speed-up over the original resampling algorithm.

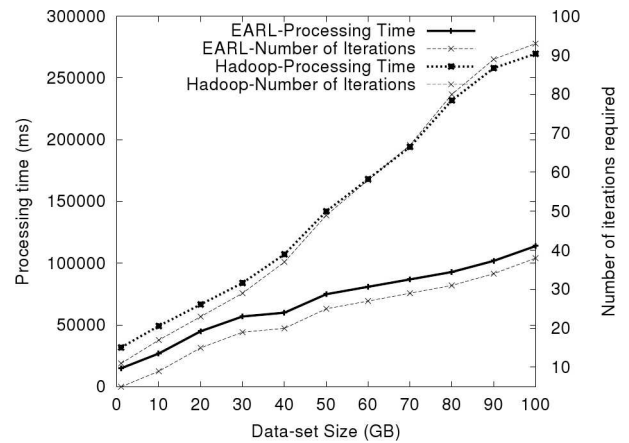

Figure 7: Computation of K-Means using EARL

### 6.3 EARL and Advanced Mining Algorithms

*EARL* can be used to provide early approximation for advanced mining algorithms, and this experiment provides a performance study when using *EARL* to approximate *K-Means*.

It is well known that *K-Means* algorithm converges to a local optima and is also sensitive to the initial centroids. For these reasons the algorithm is typically restarted from many initial positions. There are various techniques used to speed up K-Means, including parallelization [31]. Our approach, compliments previous techniques by speeding up K-Means without changing the underlying algorithm.

Figure 7 shows the results of running *K-Means* with EARL and stock Hadoop. Our approach leads to a speed up due to two reasons: (1) K-Means is executed over a small sample of the original data and (2) K-Means converges more quickly for smaller data-sets. Because of a synthetic data-set, we were also able to validate that EARL finds centroids that are within 5% of the optimal.

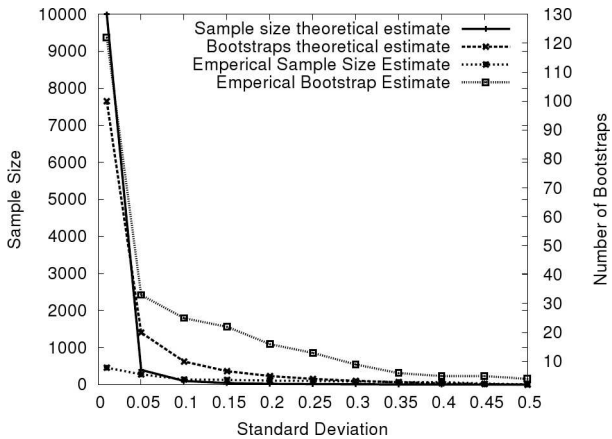

**Figure 8: Empirical sample size and number of bootstraps estimates vs. a theoretical prediction**

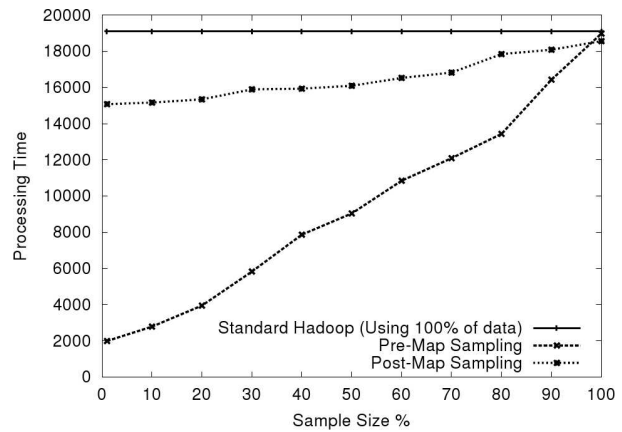

**Figure 9: Processing times of pre-map and post-map sampling**

## 6.4 Sample Size and Number of Bootstraps

In this experiment we measure how the theoretical sample size and the theoretical number of bootstraps prediction compare to our empirical technique of estimating the sample size and the number of bootstraps. We use a sample mean as the function of interest. Frequently, theoretical prediction for sample size is over estimated given a low error tolerance and is under-estimated for a relatively high error tolerance. Furthermore, theoretical bootstrap prediction frequently under-estimates the required number of bootstraps. In other empirical tests we have observed cases where theoretical bootstrap prediction is much higher than the practical requirement. This makes a clear case for the necessity of an empirical way to determine the required sample size and the number of bootstraps to deliver the user-desired error bound. In the case of the sample mean, we found that for a 5% error threshold, a 1% uniform sample and 30 bootstraps are required.

## 6.5 Pre-map and Post-map Sampling

In this experiment we determine the efficiency of pre-map and post-map sampling as described in Section 3.3 when applied to computation of the mean. Recall that pre-map sampling is done before sending any input to the mapper thus significantly decreasing the load-times and improving response time. The down-side of pre-map sampler is a potential decrease in accuracy of estimating the number of *key, value* pairs which may be required for correcting the final output. In post-map sampling, the sampling is done per-key, which increases the load-times but potentially improves accuracy of estimating the number of *key, value* pairs. As presented in Figure 9 the pre-map sampling is faster than post-map sampling in terms of total processing time. Furthermore, our empirical evidence suggests that for a large sample size, pre-map sampler is as accurate in terms of the number of *key, value* prediction as the post-map sampler. Therefore, to decrease the load-times, and to produce a reasonable estimate for functions that require result correction, the *pre-map* sampler should be used. The *post-map* sampler should be used when load-times are of low concern and a fast as well as accurate estimates of a function on a relatively small sample size are required.

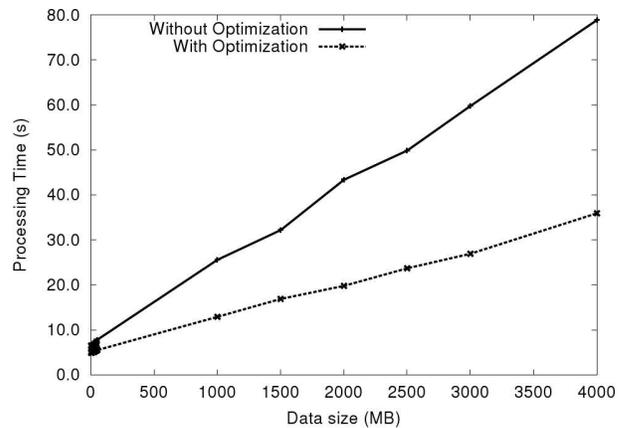

**Figure 10: Processing time with the update procedure**

## 6.6 Update Overhead

This experiment measures the benefit that our incremental processing strategies (inter-iteration and intra-iteration) achieve. Recall that in order to produce samples of larger sizes and perform resampling efficiently, we rely on delta maintenance as described in Section 4. Figure 10 shows the total processing time of computing the mean function with and without the delta maintenance optimization. The data-size represents the total data that the function was to process. The without optimization strategy refers to executing the function of interest on the entire dataset and with optimization strategy refers to execution the function on half of the data and merging the results with the previously saved state as described in Section 10. The optimized strategy clearly outperforms the non-optimized version. The optimized strategy introduced achieved a speedup of close to 300% for processing a 4GB data-set as compared to the standard method.

## 7. RELATED WORK

Sampling techniques for Hadoop were studied previously in [14] where authors introduce an approach of providing Hadoop job with an incrementally larger sample size. The authors propose an *Input Provider* which provides the *JobClient* with the initial subset of the input splits. The subset of splits used for input are chosen randomly. More splits are provided as input until the JobTracker declares that enough data had been processed as indicated by the required sample size. The claim that the resulting input represents a uniformly random sample of the original input is not well validated. Furthermore the assumption that each of the initial splits represents a random sample of the data, which would justify the claim that the overall resulting sample is a uniform sample of the original data, is not well justified. Finally the authors do not provide an error estimation framework which would make it useful for the case where only a small sample of the original data is used. Overall, however, [14] provides a very nice overview of a useful sampling framework for Hadoop that is simple to understand and use.

Random sampling over database files is closely related to random sampling over HDFS and the authors in [20] provide a very nice summary of various file sampling techniques. The technique discussed in [20] that closely resembles our sampling approach is known as a *2-file technique combined with an ARHASH method*. In the method, a set of blocks, $F_1$, are put into main memory, and the rest of the blocks, $F_2$, reside on disk. When seeking a random sample, a 2-stage sampling process is performed where $F_1$ or $F_2$ is first picked randomly, and then depending on the choice, a random sample is drawn from memory or from disk. The expected number of disk seeks under this approach is clearly much less than if only the disk was used for random sampling. The method described, however, is not directly applicable to our environment and therefore must be extended to support a distributed filesystem.

Authors in [5, 10] explore another efficient sampling approach, termed *block sampling*. Block-sampling suffers, however, from a problem that it no longer is a uniform sample of the data. The approximation error derived from a block-level sampling depends on the layout of the data on disk (i.e., the way that the tuples are arranged into blocks). When the layout is random, then the block-sample will be just as good as a uniform tuple sample, however if there is a statistical dependence between the values in a block (e.g., if the data is clustered on some attribute), the resulting statistic will be inaccurate when compared to that constructed from a uniform-random sample. In practice most data layouts fall somewhere between the clustered and the random versions [5]. Authors in [5] present a solution to this problem where the number of blocks to include in a sample, are varied to achieve a uniformly random distribution.

Approximation techniques for data-mining were also extensively studied, however the approximation techniques introduced are not general (e.g., specific to association rule mining [6]). The authors in [6] propose a two phase sampling approach, termed *FAST*, where a large sample $S$ is first taken and then a smaller sample is derived from $S$. This tailored sampling technique works well, giving speedups up to a factor of 10. Although the proposed approach is highly specialized, *EARL* can still obtain comparable results.

Hellerstein et al. [8] presented a framework for *Online Aggregation* which is able to return early approximate results when querying aggregates on large stored data sets [15]. This work is based on relational database systems, and is limited to simple single aggregations, which restricts it to `AVG`, `VAR`, and `STDDEV`. Work in [22] provides online support for large map-reduce jobs but is again limited to simple aggregates. Similarly, work by [24] provides an approximation technique using bootstrapping, however the optimization presented was only studied in the context of simple aggregates. Later, B. Li et. al [19] and Condie et al. [8] built systems on top of MapReduce to support continuous query answering. However, these systems do not provide estimation of the accuracy of the result.

*EARL* can potentially plug into other massively parallel systems such as Hyracks [3]. Hyracks is a new partitioned-parallel software platform that is roughly in the same space as the Hadoop open source platform. Hyracks is designed to run data-intensive computations on large shared-nothing clusters of computers. Hyracks allows users to express a computation as a DAG of data operators and connectors. A Possible future direction would be to use *EARL* on Hyracks' DAG.

*EARL* relies on dynamic input support in order to incrementally increase sample size as required. While Hadoop extensions such as HaLoop [4] can support incremental data processing, this support is mainly aimed at the iterative execution model of data-mining algorithms (e.g., K-Means). HaLoop dramatically improves the iterative job execution efficiency by making the task scheduler loop-aware and by adding various caching mechanisms. However, due to its batch-oriented overhead, HaLoop, is not suitable for tasks that require dynamically expanding input.

## 8. CONCLUSION

A key part of big data analytics is the need to collect, maintain and analyze enormous amounts of data efficiently. With such application needs, frameworks like MapReduce are used for processing large data-sets using a cluster of machines. Current systems however are not able to provide accurate estimates of incremental results. In this paper we presented *EARL*, a non-parametric extension of Hadoop that delivers early results with a reliable accuracy estimates. Our approach can be applied to improve efficiency of fault-tolerance techniques by avoiding the restart of failed nodes if the desired accuracy is reached. Our approach builds on re-sampling techniques from statistics. To further improve the performance, EARL supports various optimizations to the resampling methods which makes the framework even more attractive. The paper also introduced sampling techniques that are suitable for a distributed file-system. Experimental results suggest that impressive speed-ups can be achieved for a broad range of applications. The experimental results also indicate that this is a promising method for providing the interactive support that many users seek when working with large data-sets. A direction for the future is to investigate other resampling methods (e.g., jackknife) that although are not as general and as robust as bootstrapping can still provide better performance in specific situations.

**Acknowledgments**. We thank Mark Handcock for his inspiring guidance and Alexander Shkapsky for his helpful discussions. We also thank the anonymous reviewers of this paper for their valuable feedback. This work was supported in part by NSF (Grant No. IIS 1118107).

# APPENDIX

## A. CATEGORICAL AND INTER-DEPENDENT DATA

In this paper w.l.o.g., we assumed that $N$ consists of numerical data and that $f$ also returns numerically ordered results. Our approach, however, is also applicable to categorical data with a small modification discussed next.

The analysis of categorical data will involve the proportion of "successes" in a given population. The success can be defined as an estimate of the parameter of interest. Therefore, given a random sample of size $n$ the number of successes $X$ divided by the sample size gives an estimate for the population proportion $p$. This proportion follows a binomial distribution with mean $p$ and variance $\frac{p(1-p)}{n}$. Because the binomial distribution is approximately normal, for large sample sizes, a z-test can be used for testing confidence interval and significance. This approach allows *EARL* to be applied even to categorical data.

We have also assumed that all samples contain i.i.d. data, however the bootstrap technique can be modified to support non-iid (dependent) data when performing resampling [27, 17, 25, 18]. The approach used to deal with b-dependent data is usually called block-sampling. A data-set that is b-dependent contains $\frac{N}{b}$ blocks where each block $X_i, ..., X_{i+b}$ represents $b$ inter-dependent tuples. Such dependency is usually present in time-series data. In the block based sampling instead of a single observation, blocks of consecutive observations are selected. Such a sampling method insures that dependencies are preserved amongst data-items. In this work we focus on i.i.d. data nevertheless, as stated, inter-dependent data can also be supported.